\documentclass[11pt]{article}
\usepackage[margin=1in]{geometry}
\usepackage{array}
\usepackage{amssymb, bbm}
\usepackage{mathbbol}
\usepackage{amsmath}
\usepackage{natbib}
\usepackage{graphicx}
\usepackage{booktabs}
\usepackage{multirow}
\begin{document}
\title{{\sc Reduced Form Capital Optimization}}
\newcommand{\ind}{\mathbbm{1}}
\newcommand{\cov}{\textup{cov}} 
\newcommand{\var}{\textup{var}}

\author{Yadong Li, Dimitri Offengenden and Jan Burgy\thanks{Barclays, we thank Oksana Kitaychik, Art Mbanefo, Barry McQuaid, Jeff Nisen, Ariye Shater and Yishen Song for many useful discussions and comments. The views expressed in the paper are the authors' own, they do not necessarily represent the views of Barclays.} }
\maketitle

\abstract
We formulate banks' capital optimization problem as a classic mean variance optimization, by leveraging an accurate linear approximation to the Shapely or Constrained Aumann-Shapley (CAS) allocation of max or nested max cost functions. This reduced form formulation admits an analytical solution, to the optimal leveraged balance sheet (LBS) and risk weighted assets (RWA) target of banks' business units for achieving the best return on capital.

\section{\label{intro}Introduction}

The regulatory capital requirement for banks has been on a rising trend since the 2008 financial crisis. Given the increasing cost in capital, banks are strongly incentivized to optimize their capital utilization among their business units to achieve better return on capital (RoC).  Today, improving capital efficiency is the center piece of a bank's business strategy.

Despite being a primary concern for banks' leadership teams, a rigorous and practical quantitative method has not been established for capital optimization. The main challenge in capital optimization is its enormous complexity, as it depends on a bank's entire business operation: including trading, lending, revenue, expenses and management strategies. Among those, the regulatory capital  calculation is one of the most complex aspects.

A global bank often sets up multiple subsidiary legal entities, in order to compete effectively in local markets. Each of the legal entities is regulated by their respective regional banking authorities. In addition, a bank has to report its consolidated balance sheet and capital for its top level legal entity, which is subject to the rules and regulations of its main domestic regulator. A bank's overall capital is therefore the greater of 1) the consolidated capital of its top legal entity, or 2) the sum of all subsidiary legal entities' capital. 

The regulatory capital of a single legal entity is the greater of the required capital for risk weighted assets (RWA) and leverage balance sheet (LBS), as described in \cite{b3}:
\begin{equation}
\label{cap}
\textup{Legal Entity Capital} \ge \max(\textup{CET1}\times\textup{RWA}, \textup{T1}\times\textup{LBS})
\end{equation}
where the CET1 is the core tier one capital ratio for RWA, and the T1 is the tier one capital ratio for LBS\footnote{There is a slight difference between the capital structure corresponding to the CET1 and T1 ratios, as hinted by their names; the difference lies in the additional tier one (AT1) instruments, such as Coco bonds. This difference can be compensated by adjusting the T1 ratio.}. Subsequently, we loosely refer to the $\textup{CET1}\times\textup{RWA}$ and $\textup{T1}\times\textup{LBS}$ as the RWA and LBS capital. Under Basel 3, the minimum CET1 ratio is 4.5\% and the minimum T1 ratio is 3\%. However, banks usually set up internal management CET1 and T1 target ratios well above their regulatory minima, the internal management targets factor in stress capital and management buffers. The LBS and RWA in \eqref{cap} are extremely complicated nonlinear functions depending on many underlying factors, such as a bank's trading position, balance sheet, market risk factors and even historical data. 

One possible approach for capital optimization is to build a structural model for a bank's entire operation, similar to the CCAR (\cite{ccar18}). Such a structural model would have to be detailed enough so that bank's revenue and capital can be related to various underlying driving factors. With this approach, an optimal dynamic strategy could be found from the structural model by varying the driving factors. \cite{bis} is a recent attempt at the structural model approach; it derives the optimal capital strategy of a stylized bank with two business units over three time periods.
The difficulty of such structural model approach is that it quickly becomes very complicated and intractable; and it incurs high level of model risk because of the large number of model assumptions and the likelihood of missing or mis-specifying important factors and relationships. 

Instead, we take a much simpler and more practical reduced form approach, by expressing a bank's optimal capital strategy in terms of allocated capital. The allocated capital is the result from a strict additive allocation of a bank's overall capital to individual business units. Readers are referred to \cite{cas} for a detailed description of various capital allocation methods and their properties. A business unit's allocated capital is usually smaller than its standalone capital, because the allocated capital takes into account hedging and diversification benefits. One of the main usages of allocated capital is to compute the RoC of individual business units, so that the hedging and diversification effects are included for fair performance comparisons.

By treating allocated capital as exogenous factors, we greatly simplified the formulation of capital optimization because we no longer need to model any of the underlying driving factors or the complicated business operations of a bank; as its capital is simply the sum of the allocated capital by construction. It is a routine practice for a bank's senior management to set limits or targets on individual business units' allocated capital. Thus, an optimal capital strategy expressed in terms of allocated capital can be easily implemented by changing the corresponding targets on allocated capital, and relying upon the well-established management structure and protocol to achieve them.

The $\max$ function in \eqref{cap} poses another challenge: it renders the nonbinding capital component (which is the smaller of the RWA and LBS capital) irrelevant for a legal entity's overall capital, making it very difficult for managers to understand and reason about the impact of the nonbinding capital component. As a result, it is common for banks to only actively manage the binding capital constraint, which is easier but clearly insufficient and sub-optimal.

It is straightforward to conclude that a bank can only be optimal in RoC when its consolidated top level RWA and LBS capital are equal. Otherwise there is either ``free''  RWA or ``free'' LBS that could be used to generate additional revenue without increasing a bank's capital. However, in practice there are various business constraints that prevent a bank from operating at its global optimal RoC with equal RWA and LBS capital. For example, it may not be feasible for a bank to dramatically change the footprint of business units that caused the imbalance of its RWA and LBS capital. As these practical business constraints are difficult to model quantitatively, we take a local optimization approach, which is to find a local optimum of RoC near a bank's current capital positions, instead of finding the global optimum in RoC. The local optimum is well suited for a bank's day to day capital management as it would not be difficult for a bank to implement small changes from its current capital position.

Another challenge of capital optimization is to ensure consistency in optimizations across a bank's business unit hierarchy. In practice, capital optimization is performed at multiple levels of a bank, where each business unit may take independent actions to optimize their own RoCs. As previously mentioned, a business unit's RoC depends on capital allocation method for its denominator, therefore different capital allocation methods will result in different RoCs, leading to different incentives for business units to optimize against. As a result, it is important for a capital allocation method to set up the correct business incentives so that when individual business units attempt to optimize their own RoC it also results in an improvement in a bank's overall RoC. The business incentives of different allocation methods were studied in \cite{cas}, showing that the Shapley or Constrained Aumann-Shapley (CAS) allocations give the best business incentives among common allocation methods. 

To illustrate business incentives afforded by different allocation methods, we consider a stylized example of a bank with five business units in Table \ref{tab1}. In this example, we assume the bank's total LBS and RWA capital are simple sums of those of individual business units, i.e. we ignored the diversification benefits between different business units in either LBS or RWA. However, there remains a significant diversification benefit between the bank's RWA and LBS capital because of the nonlinear max function\footnote{The sum of individual business units' standalone capital is 1180, and the bank's total capital is 1000; thus, a diversification equals $1180 - 1000 = 180$.}. In the table, we show the results from different allocation methods, such as Euler, Standalone (S/A in the table) and Shapley/CAS allocations, and the corresponding RoCs under a simple assumption for revenue: A business unit's revenue is 10\% of its standalone capital. The ``Linear'' column under ``Allocation'' in Table \ref{tab1} is a linear approximation to the Shapley allocation, which we will explain in detail later.

The example in Table \ref{tab1} shows that the Shapley/CAS\footnote{In this particular example, the CAS allocation is identical to the Shapley allocation.} allocation indeed gives a more sensible allocation and better business incentives than other allocation methods. For example, it recognizes that the firm's current binding capital constraint is in the LBS, therefore D receives a smaller allocation than B or C (even though the three have identical standalone capital). C also receives more allocation than B because of more RWA capital. In contrast, the Standalone allocation offers no incentive for closing the RWA and LBS gap by giving identical allocations to C and D. The Euler allocation goes to the opposite extreme of totally disregarding the RWA capital, and giving identical allocation to B and C, despite C's larger RWA capital. Overall, the Shapley allocation produces more balanced ``relative costs'' of the RWA and LBS capital, which not only strongly incentivizes the business units to close the gap between RWA and LBS capital, but also rewards them for reducing their non-binding RWA capital. As a result, the RoC computed from Shapley allocation in Table \ref{tab1} is a much better measure of the relative performance of the business units than the RoCs computed from other allocation methods.

\begin{table}[]
\centering
\caption{A Stylized Example of LBS vs RWA Capital}
\label{tab1}
\vspace{.2cm}

\scriptsize
\begin{tabular}{c|rr|r|rrrr|rrrr}
\hline
\multirow{2}{*}{\textbf{\begin{tabular}[c]{@{}c@{}}Business\\ Unit\end{tabular}}} & \multicolumn{1}{c}{\multirow{2}{*}{\textbf{\begin{tabular}[c]{@{}c@{}}RWA\\ Capital\end{tabular}}}} & \multicolumn{1}{c|}{\multirow{2}{*}{\textbf{\begin{tabular}[c]{@{}c@{}}LBS\\ Capital\end{tabular}}}} & \multicolumn{1}{l|}{\multirow{2}{*}{\textbf{Revenue}}} & \multicolumn{4}{c|}{\textbf{Allocation}}                                                                                                                   & \multicolumn{4}{c}{\textbf{RoC}}                                                                                                                    \\ \cline{5-12} 
                                                                                  & \multicolumn{1}{c}{}                                                                                & \multicolumn{1}{c|}{}                                                                                & \multicolumn{1}{l|}{}                                  & \multicolumn{1}{l}{\textbf{S/A}} & \multicolumn{1}{c}{\textbf{Euler}} & \multicolumn{1}{l}{\textbf{Shapley}} & \multicolumn{1}{l|}{\textbf{Linear}} & \multicolumn{1}{c}{\textbf{S/A}} & \multicolumn{1}{c}{\textbf{Euler}} & \multicolumn{1}{c}{\textbf{Shapley}} & \multicolumn{1}{c}{\textbf{Linear}} \\ \hline
A                                                                                 & 230                                                                                                & 150                                                                                                 & 23                                                    & 195                                    & 150                               & 179                                 & 179                                 & 0.118                                   & 0.153                              & 0.128                                & 0.129                               \\
B                                                                                 & 120                                                                                                & 250                                                                                                 & 25                                                    & 212                                    & 250                               & 218                                 & 218                                 & 0.118                                   & 0.1                                & 0.115                                & 0.115                               \\
C                                                                                 & 150                                                                                                & 250                                                                                                 & 25                                                    & 212                                    & 250                               & 228                                 & 227                                 & 0.118                                   & 0.1                                & 0.11                                 & 0.11                                \\
D                                                                                 & 250                                                                                                & 150                                                                                                 & 25                                                    & 212                                    & 150                               & 186                                 & 185                                 & 0.118                                   & 0.167                              & 0.134                                & 0.135                               \\
E                                                                                 & 150                                                                                                & 200                                                                                                 & 20                                                    & 169                                    & 200                               & 188                                 & 191                                 & 0.118                                   & 0.1                                & 0.106                                & 0.105                               \\ \hline
\textbf{Total}                                                                    & \textbf{900}                                                                                        & \textbf{1000}                                                                                        & \textbf{118}                                           & \textbf{1000}                           & \textbf{1000}                      & \textbf{1000}                        & \textbf{1000}                        & \multicolumn{4}{c}{\textbf{0.118}}                                                                                                                       \\ \hline
\end{tabular}
\end{table}

In the rest of this article, we first present an important linear approximation to Shapley/CAS allocation of the cost function in the format of \eqref{cap}. We then use this approximation to formulate the reduced form capital optimization.
 
\section{Shapley Allocation of the Max}

Shapley allocation specifies that the allocation to a business unit is the average of its incremental contributions over all possible permutations of the business units. We use $\Omega$ to represent the full set of business units in a bank, each being indexed and identified by an integer from 1 to $n$.  Here the term ``business unit'' is used in a very generic sense, it could mean a business division, a trading desk, a trading book or even an individual trade. We use $\tilde{\pi}$ to represent a random permutation across all the business units, and $\tilde{\pi}(k)$ is the position of business unit $k$ in the permutation $\tilde{\pi}$; therefore, the $S(k; \tilde{\pi}) = (i; \forall \tilde{\pi}(i) < \tilde{\pi}(k))$ is the set of business units positioned ahead of the business unit $k$ for the given permutation $\tilde{\pi}$.
 
Using this notation, the Shapley allocation to the business unit $k$ can be written as:
\[
\alpha_k = \frac{1}{n!}\sum_{\tilde{\pi}}\left(c(S(k; \tilde{\pi}) \cup k) - c(S(k; \tilde{\pi})\right)
\]
where $c(S)$ is a cost function for a given subset $S \subset \Omega$ of business units and the average is over all possible $n!$ permutations of the $n$ business units.

\subsection{\label{lin}Shapley Allocation of $\max(\sum_{i \in \Omega} a_i, \sum_{i \in \Omega} b_i)$}

We consider the following allocation problem: each business unit $i$ in $\Omega$ has two associated cost metrics $a_i$ and $b_i$, e.g., LBS capital and RWA capital. These two cost metrics are additive and the overall cost of any subset of   $S \subset \Omega$ is the greater of the two, i.e.:  $c(S) = \max(\sum_{i \in S} a_i, \sum_{i \in S} b_i)$. The total cost to be allocated is therefore $c(\Omega) = \max(\sum_{i \in \Omega} a_i, \sum_{i \in \Omega} b_i)$. 

In the cost function $c(S) = \max(\sum_{i \in S} a_i, \sum_{i \in S} b_i)$, the arguments in $\max(\cdot)$ are linear in $a_i$ and $b_i$. Therefore the incremental contribution of business unit $k$ in an arbitrary permutation $\tilde{\pi}$ is its add-on to the dominant side of the running sum $\sum_{i \in S} a_i $ or $\sum_{i \in S} b_i $, where $S$ is the set of business units ahead of $k$ in $\tilde{\pi}$; for now we ignore the case that the dominant side may switch by adding the unit $k$ itself to $S$. Therefore, if we use $p(a < b; k)$ to represent the probability that the running sum up to $k$ in a random permutation $\tilde{\pi}$ is greater for the sum of $b$,
then we have a simple linear approximation of the average incremental contribution of unit $k$, a.k.a the Shapley allocation to the unit $k$: 
$\alpha_k \approx (1- p(a < b; k)) a_k + p(a < b; k) b_k$.

We now consider how to compute the $p(a < b; k)$:
\begin{eqnarray}
\label{appp}
\nonumber
p(a < b; k) &=& \mathbb P[\sum_{i \in S} a_i < \sum_{i \in S} b_i] = \mathbb P[\sum_{i \in S} (a_i - b_i) < 0] \\
&=& \mathbb P[\sum_i \ind_{i \in S} (a_i - b_i) < 0]
\end{eqnarray}
The indicator $\ind_{i\in S}$ represents whether the unit $i$ is ahead of $k$ in a random permutation $\tilde{\pi}$. Even though the business unit $k$ itself is not part of $S$ when considering the allocation to $k$ itself, we choose to include $\ind_k (a_k - b_k)$ in the summation in \eqref{appp} for a better approximation to $p(a<b; k)$. By including $\ind_k (a_k - b_k)$, we effectively used the average of $\frac{1}{2}(a_k+b_k)$ as $k$'s incremental contribution in case adding $k$ changes the dominant side of the running sums\footnote{This can be easily shown by enumerating all 4 possible combinations of the dominant side before and after adding $k$.}. This is a very good approximation because the allocation to the element $k$ is strictly between $a_k$ and $b_k$ under such circumstances. The other advantage of including the $\ind_k (a_k - b_k)$ is that the $p(a<b; k)$ no longer depends on $k$, thus simplifying the formulation. Thus we subsequently denote it as $p(a<b)$.

The $p(a < b)$ can be approximated analytically by moment matching the $\tilde s = \sum_i \ind_{i \in S} (a_i - b_i)$ using a normal distribution, whose mean and variance are known analytically (see appendix \ref{appa}): 
\begin{eqnarray}\nonumber
\mu_s &=& \frac{1}{2} \sum_i (a_i - b_i) \\ 
\label{p}
\sigma_s^2 &=& \frac{1}{6}\sum_i \left(a_i - b_i\right)^2 + \frac{1}{12} \left(\sum_i (a_i - b_i)\right)^2 \\
\nonumber
p(a < b) &\approx& \Phi(-\mu_s/\sigma_s)
\end{eqnarray}
where $\Phi(\cdot)$ is the normal distribution function. Strictly speaking, the $\tilde s$ is not normal because the central limit theorem does not apply as $\ind_{i \in S} (a_i - b_i)$ are correlated with correlation of $\frac{1}{3}$ (see appendix \ref{appa}). Further adjustments to higher moments could be made to account for the deviation from normality, potentially making the moment matching more accurate. However, we found that the normal moment matching is already quite accurate in practice, thus there is no need for such adjustments.

\begin{figure}
\caption{\label{fig:a1}Shapley Allocation of $\max(\sum_i a_i, \sum_i b_i)$: Correlation(MC, \eqref{a1})}

\center

\includegraphics[width=0.6\linewidth]{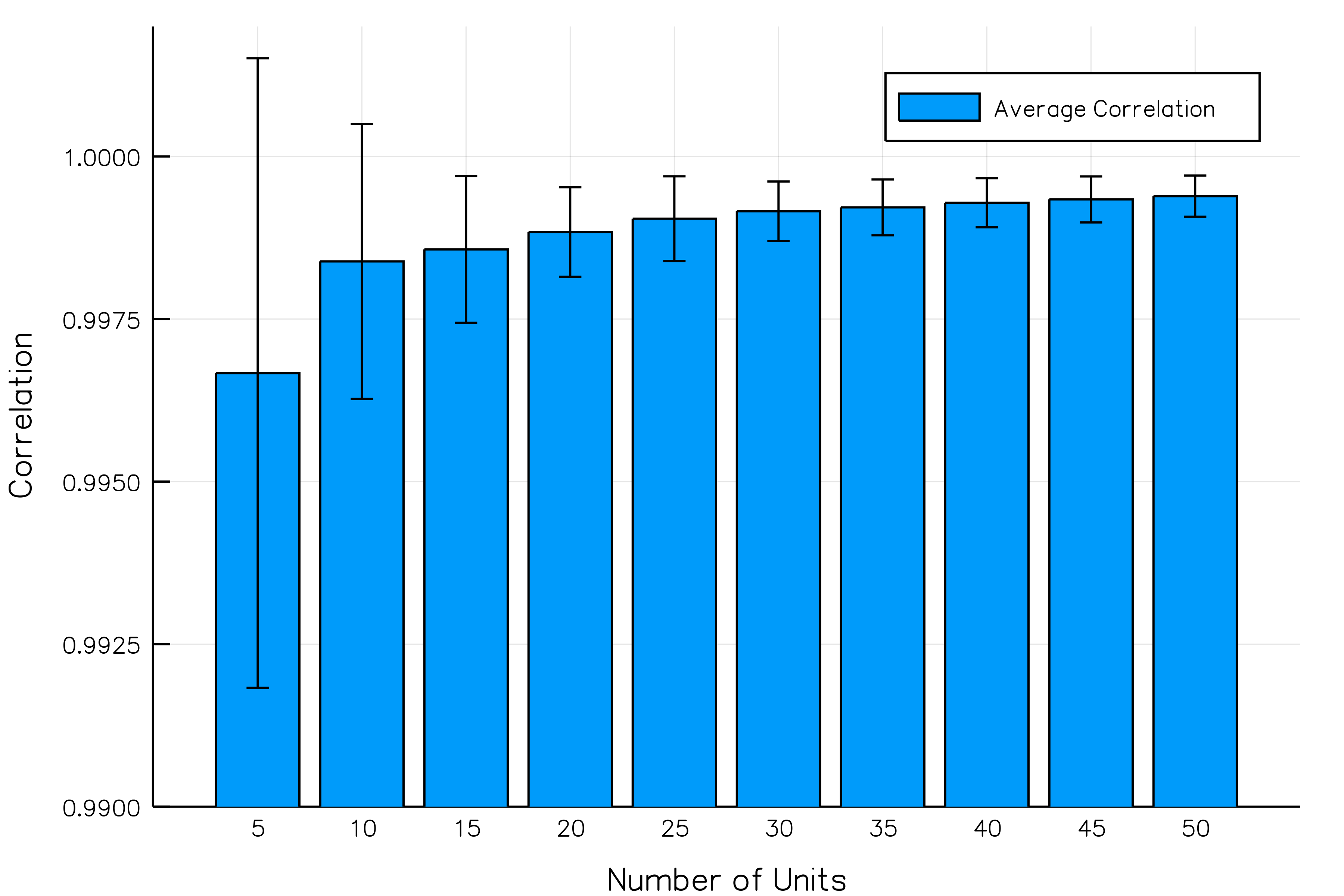}

\end{figure}

Figure \ref{fig:a1} is a numerical test of the linear approximation to the Shapley allocation using \eqref{p}: we varied the number of business units $n$ from 5 to 50, for each $n$ we  drew 20 sets of $a_i$s and $b_i$s from a uniform distribution, and computed the correlation between the exact Shapley allocation results from Monte Carlo (MC) simulation and the linear approximation from \eqref{p}. Figure \ref{fig:a1} is the average and standard deviation of the correlations from the 20 random samples for each $n$, which shows that the approximation \eqref{p} is quite accurate, even for $n$ as small as 5. The correlation between the numerical solution and the linear approximation is greater than 99.5\% for all $n$, and the accuracy improves with larger $n$, the correlation goes above 99.9\% for $n>20$.

Although the linear approximation comes quite close, the linear approximation does not add up to the exact total cost of $c(\Omega)$, therefore, we introduce a correction factor $\beta$ to ensure the additivity:
\begin{equation}
\label{a1}
\alpha_k = \beta\left(\left(1-p(a < b)\right)a_k + p(a < b) b_k\right)
\end{equation}
\begin{equation}
\label{alpha}
\beta = \frac{\max\left(\sum_i a_i, \sum_i b_i\right)}{\sum_i a_i - 2 p(a<b) \mu_s }.
\end{equation}
where the $\mu_s$ and $p(a<b)$ are given in \eqref{p}. In Table \ref{tab1}, we also show that the linear approximation result from \eqref{a1} is very close to the corresponding Shapley/CAS allocation. 

\subsection{Shapley Allocation of $\max\left(f(\Omega), g(\Omega)\right)$}
We now consider the approximation to the Shapley allocation of the more generic cost function $c(S) = \max\left(f(S), g(S)\right)$, which is exactly \eqref{cap} if $f(S), g(S)$ are the LBS and RWA capital of a subset $S \subset \Omega$ of business units.

We denote the Shapley allocation of $f(\Omega)$ and $g(\Omega)$ to business unit $k$ as $\alpha_k^f$ and $\alpha_k^g$ respectively so that $f(\Omega)=\sum_{i \in \Omega} \alpha_i^f$ and $g(\Omega)=\sum_{i \in \Omega} \alpha_i^g$ by construction. In practice, banks do compute separately the RWA and LBS allocation to its business units. Obviously, the Shapley allocation of $\max\left(f(\Omega), g(\Omega)\right)$ is not the max of their respective allocations because the max function is nonlinear, i.e., $\alpha_k \ne \max(\alpha_k^f, \alpha_k^g)$. 

We propose a simple approximation to the cost function in the form of $c(S) = \max\left(f(S), g(S)\right)$. The key idea is to approximate $f(S)$ and $g(S)$ by their respective Shapley allocations, so that $c(S) \approx l(S) = \max\left(\sum_{i \in S} \alpha_i^f, \sum_{i \in S} \alpha_i^g\right)$. By construction, $c(\Omega) = l(\Omega)$ for the entire portfolio $\Omega$ because $f(\Omega) = \sum_{i \in \Omega} \alpha_i^f$ and $g(\Omega) = \sum_{i \in \Omega} \alpha_i^g$. The $c(S)$ and $l(S)$ track each other closely for any subsets $S \in \Omega$, as after all the Shapley allocation $\alpha^f$ and $\alpha^g$ are the expected incremental contributions to $f(S)$ and $g(S)$. In essence, we replaced the random incremental contributions to $f(S)$ and $g(S)$ by their respective expectations of $\alpha^f$ and $\alpha^g$.

\begin{figure}
\caption{\label{fig:a2}Correlation($c(S)$, $l(S)$) for VaR}
\center
\includegraphics[width=.5\textwidth]{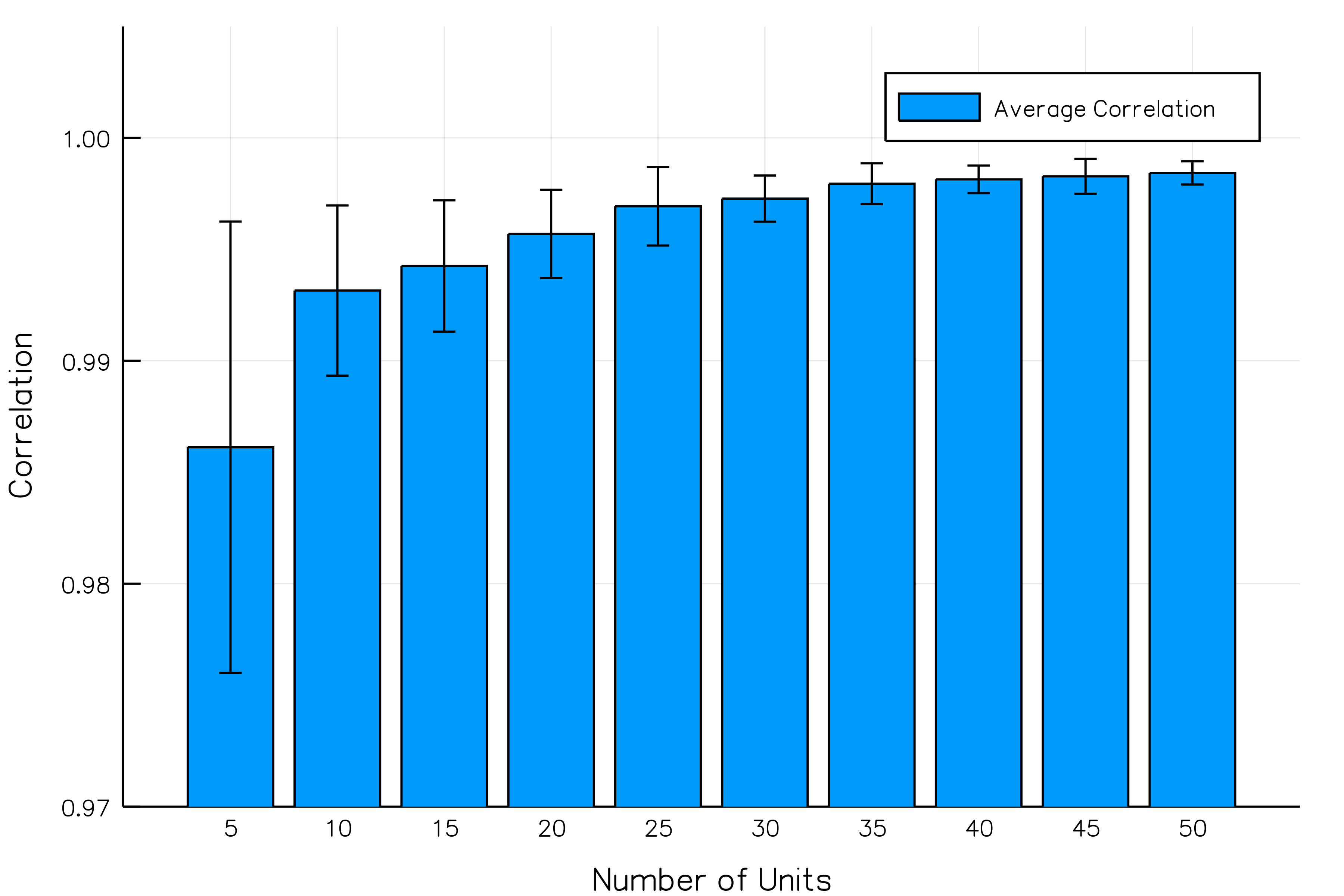}
\end{figure}

Figure \ref{fig:a2} is a numerical example of 99\% VaR and its linear approximation. We generated 20 sets of random PnL vectors for each of the $n$ business units with randomized PnL volatility and correlations, then compared the mean and standard deviation of the 20 resulting correlations between $f(S) = \textup{VaR}(S)$ and its linear approximation $\sum_{i \in S} \alpha_i^f$ under many random permutations. It shows that the linear approximation to the cost function is very accurate; the average correlation between the actual and linearized cost function is greater than 99.5\% when $n>20$. The correlation is greater than 98\% even for $n$ as small as 5.

We do expect this high level of accuracy from the linearized cost function $l(S)$ in practice because most risk and capital measures, such as VaR, do exhibit linear-ish behaviors at the top level business units within legal entities. The reason is that these top level business units tend to have complicated and diverse portfolios, thus their risk and capital metrics are unlikely to exhibit strong non-linear behaviors.

The Shapley allocation of $c(S)$ can then be well approximated by that of $l(S)$, which is given by \eqref{a1}. Therefore:
\begin{equation}
\label{a2}
\alpha_k \approx \beta \left(1-p\left(\alpha^f < \alpha^g\right)\right)\alpha_k^f + \beta p\left(\alpha^f < \alpha^g\right) \alpha_k^g
\end{equation}
where $p\left(\alpha^f < \alpha^g\right)$ is given in \eqref{p} and $\beta$ is defined in \eqref{alpha}.

Similar to Shapley allocation, the CAS allocation is linear with respect to its cost function, therefore the same linear approximation \eqref{a2} also holds true under CAS for all the business units in the same legal entity. The simple linear relationship between the RWA allocation, LBS allocation and capital allocation of a business unit in \eqref{a2} offers a simple, intuitive and consistent way to understand and reason about both the LBS and RWA consumptions; it overcomes the difficulty of the $\max$ function in \eqref{cap} and allows a bank to effectively and consistently manage and optimize both the binding and non-binding capital components.

It is a significant and useful result that the Shapley/CAS allocation of a complicated nonlinear cost function $c(S)=\max\left(f(S), g(S)\right)$ can be well approximated by a simple weighted average of $\alpha^f, \alpha^g$ in \eqref{a2}. This property is critical for the formulation of the reduced form capital optimization in the next section. 

\subsection{\label{le}Legal Entity Hierarchies}
It is a common practice for global banks to set up multi-level legal entity structures, in order to meet the regulatory and business requirements of regional markets. 
The linear approximation can be extended to more complicated legal entity hierarchies. However, the CAS allocation has to be used in such cases to remain consistent across the organizational structure, as explained in \cite{cas}.

Let's consider a realistic example of a bank with two non-overlapping subsidiary legal entities, $X$ and $Y$; each of them are regulated by a regional regulator with different minimal capital ratios, and the bank's consolidated legal entity with the whole portfolio of $\Omega = X \cup Y$ regulated by its domestic regulator. For a subset of business unit $S \subset \Omega$, we can write its cost function as:
\begin{equation}
\label{c6}
c(S) = \max\left(\max\left(f^\theta(S), g^\theta(S)\right), \max\left(f^x(S \cap X), g^x(S \cap X)\right) + \max\left(f^y(S \cap Y\right), g^y(S \cap Y))\right)
\end{equation}
i.e, the bank's capital is the greater of 1) the consolidated entity's capital and 2) the sum of two subsidiaries' capital. Within each legal entity, the capital is the greater of LBS and RWA capital, which are represented by $f(\cdot)$ and $g(\cdot)$; the superscripts $\theta, x$ and $y$ represent the consolidated and subsidiary legal entities. X and Y are non-overlapping: each of the top level business unit of the bank belongs to either X or Y, but not both, i.e., $X \cap Y = \emptyset $.

Following the same approach described in the previous section, we approximate the $f(S), g(S)$ by their CAS allocations:
\begin{eqnarray}
\label{abc}
l(S) &=& \max(\theta(S), x(S) + y(S)) \\ \nonumber
&=& \max\left(\max\left(\sum_{i\in S}\alpha^{f_\theta}_i, \sum_{i \in S}\alpha^{g_\theta}_i\right), \max\left(\sum_{i \in S \cap X} \alpha^{f_x}_i, \sum_{i \in S \cap X}\alpha^{g_x}_i\right) + \max\left(\sum_{i \in S \cap Y} \alpha^{f_y}_i, \sum_{i \in S \cap Y}\alpha_i^{g_y}\right)\right)
\end{eqnarray}
where the $\alpha_i^f$s and $\alpha_i^g$s are the CAS allocation of respective capital component and legal entity. $\theta(S)$, $x(S)$, $y(S)$ are defined to be the corresponding terms to ease the notation. The $c(\Omega) = l(\Omega)$ still holds for the bank's entire portfolio by construction.

Following the same logic and notation as used in section \ref{lin}, the Shapley allocation of business unit $k$ to the cost function $l(S)$ can be approximated as:
\begin{eqnarray}
\label{xi6}
\alpha_k &\approx& p(\theta > x+y, \alpha^{f_\theta} > \alpha^{g_\theta})\alpha_k^{f_\theta} + p(\theta > x+y, \alpha^{f_\theta} < \alpha^{g_\theta})\alpha_k^{g_\theta} \\ \nonumber &+& p(\theta < x+y, \alpha^{f_x} > \alpha^{g_x}) \alpha_k^{f_x} + p(\theta < x+y, \alpha^{f_x} < \alpha^{g_x}) \alpha_k^{g_x} \\ \nonumber 
&+& p(\theta < x+y, \alpha^{f_y} > \alpha^{g_y}) \alpha_k^{f_y} + p(\theta < x+y, \alpha^{f_y} < \alpha^{g_y}) \alpha_k^{g_y}
\end{eqnarray}
where all the $p(\cdot, \cdot)$s are joint probabilities of the relative ordering of respective running sums up to $k$ in a random permutation. Note that the $\alpha_k^f, \alpha_k^g$ for $X$ or $Y$ is zero if the business unit $k$ is not part of the corresponding regional legal entity. 

The joint probabilities in \eqref{xi6} can be computed very efficiently by running a small scale Monte Carlo simulation using the CAS allocation of the few top level business units. We can also define a scaling factor $\beta$ similar to the one in \eqref{alpha} to ensure the additivity of $c(\Omega) = \beta \sum_{i \in \Omega} \alpha_i$. Then the $w_j = \beta p_j(\cdot, \cdot)$, where $p_j(\cdot, \cdot)$s are the individual joint probabilities in \eqref{xi6}, can be interpreted as the ``exchange rates'' from the corresponding $\alpha_k^f, \alpha_k^g$ to the business unit's CAS capital allocation $\alpha_k$. The same set of ``exchange rates'' apply to all the business units in the bank, making it convenient to actively manage and optimize all the capital components from all legal entities.

\section{\label{co}Capital Optimization}
As discussed in the beginning, the Shapley/CAS allocations do set up the correct incentives for individual business units to take actions that improve a bank's overall RoC. A key benefit of the linear approximation is that it preserves and accentuates the correct business incentive from the Shapley/CAS allocations.

\begin{figure}
\centering
\caption{\label{fig2}``Exchange Rates'' for RWA and LBS Capital}
\vspace{.25cm}
\includegraphics[width=0.6\linewidth]{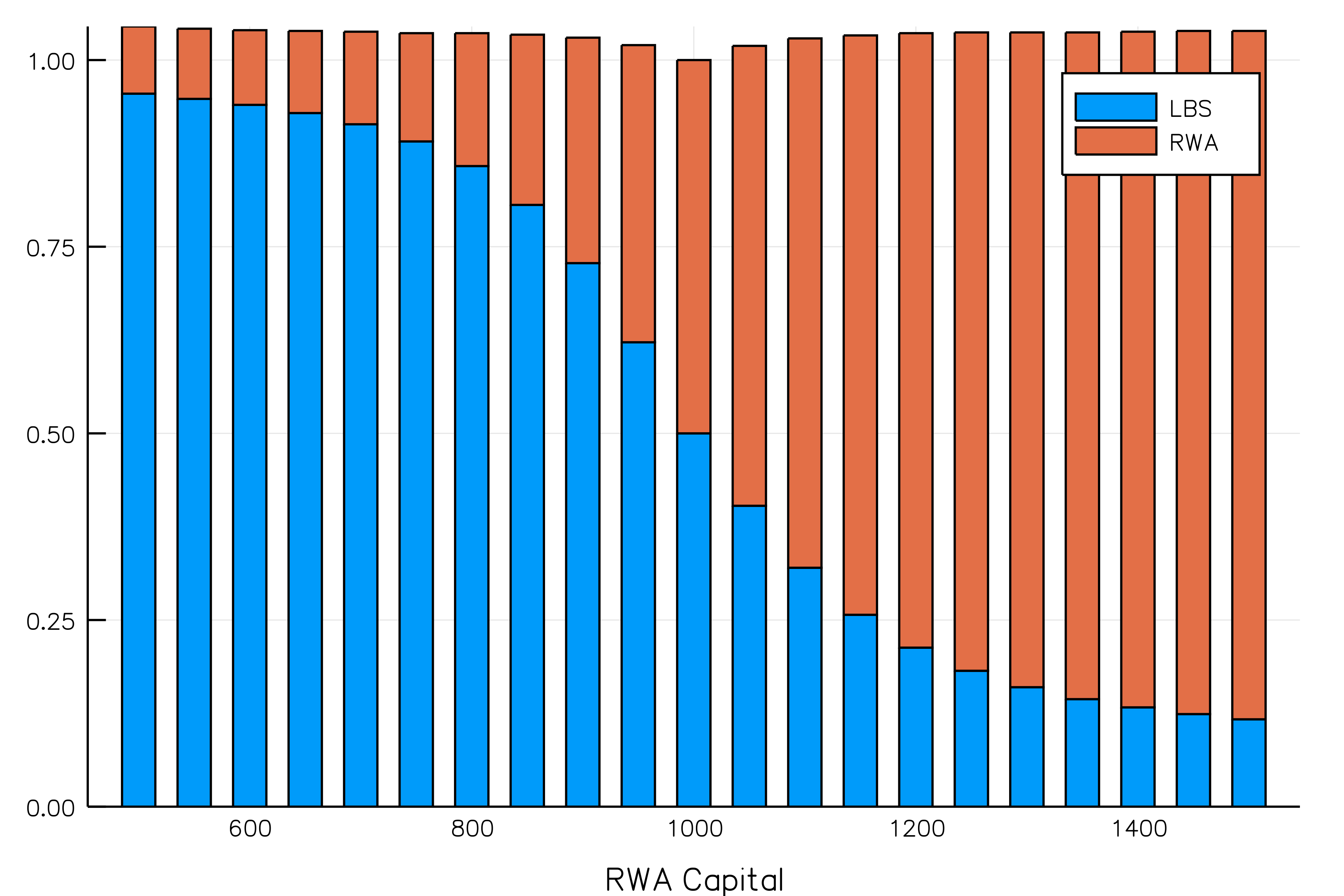}

\end{figure}

Figure \ref{fig2} is a scenario analysis using the same data as in Table \ref{tab1}, but after scaling the RWA capital of individual business units so that the RWA capital of the bank varies from 500 to 1500, while holding LBS capital constant at 1000. The color bars in the figure are the corresponding ``exchange rates'' of the RWA and LBS capital for a given scenario. Figure \ref{fig2} shows that when the bank is operating in a suboptimal state with unequal RWA and LBS capital, the ``exchange rate'' on the dominant constraint is greater, thus incentivizing the utilization of the non-binding constraint. The ``exchange rates'' are more skewed against the dominant constraint when there is a wider gap between RWA and LBS capital, providing a stronger incentive to close the gap. When the bank is at its optimal state with equal RWA and LBS capital, the ``exchange rates'' become equal at $\frac{1}{2}$, giving no incentive to move away from the optimal state.

In addition, since the linear ``exchange rates'' established a single and consistent conversion between the allocated LBS, allocated RWA and the allocated capital across all business units and all legal entities, it allows us to derive a convenient analytical solution to local capital optimization if we further assume a linear revenue model. 

We use the vector $\vec h$ to represent the allocated LBS and RWA capital of banks' $n$ business units. For the example we considered in section \ref{le}, the $\vec h$ would be a vector of length $6n$ including all six $\alpha^f, \alpha^g$ elements on the RHS of \eqref{xi6}; $\vec h$ should not be confused with $\vec \alpha$, which is a vector of length $n$ including only the LHS of \eqref{xi6}. There are two main reasons to formulate the capital optimization using the more granular $\vec h$ rather than $\vec \alpha$. The first is to be able to solve the optimal LBS and RWA allocation of every individual business unit, not just its overall capital allocation; the second is that the legal entity's capital ratio constraints are specified on its LBS and RWA, which can be expressed in $\vec h$ but not in $\vec \alpha$.

We then use $\vec r$ to represent the RoC vector for the corresponding allocated LBS and RWA capital, so that the dot product $\vec r \cdot \vec h$ is the total revenue of the bank. Similarly, we use $\vec w$ to represent the corresponding ``exchange rates'' from the linear approximation presented in the previous section, so that the dot product $\vec w \cdot \vec h$ is the sum of the allocated capital to all business units, which is exactly the bank's total capital. The $\vec w$ is a function of $\vec h$, and the RoC vector $\vec r$ is assumed to be constant. 

By linearizing both the revenue and capital, we can write down a reduced form local capital optimization similar to the classic mean variance portfolio optimization:
\begin{eqnarray}
\label{optim}
\vec \delta^* &=& \textup{argmin}_{\vec \delta} \left( \vec w^T (\vec h + \vec \delta) + \frac{\epsilon}{2} \vec \delta^T V^{-1} \vec \delta \right) \\
\textup{subject to:}& & \vec r^T \vec \delta = z, \nonumber
\end{eqnarray}
i.e., subject to a constant change $z$ in the bank's revenue, find a change $\vec \delta$ from the current allocated capital $\vec h$ so that the resulting capital position $\vec h + \vec \delta$ minimizes the firm's overall capital. The $V$ is a constant covariance matrix of $\vec h$'s daily changes, which can be conveniently estimated from historical time series of $\vec h$ following the spirit of the reduced form modeling. The covariance matrix $V$ captures the ``modes'' of the co-movements among capital components, e.g. the LBS and RWA capital changes of the same business unit are often highly correlated. The $\sqrt{\delta^T V^{-1} \vec \delta}$ is a Mahalanobis distance, which is the magnitude of $\vec \delta$ measured in the unit of the standard deviation of $\vec \alpha$'s changes. Mahalanobis distance can also be interpreted as a measure of plausibility (\cite{md}): the greater the $\sqrt{\delta^T V^{-1} \vec \delta}$, the less likely $\vec \delta$ can occur under the given covariance matrix $V$. The $\epsilon > 0$ is a penalty factor against implausible $\vec \delta$ solutions, which also controls the trade-off between capital efficiency and plausibility. 

Interestingly, there is no need to add explicit minimum ratio constraints of various legal entities to \eqref{optim}, as the optimization is expressed in the allocated capital $\vec h$ and the effects of capital ratios are implicitly captured by the RoC vector $\vec r$. Other business constraints can be easily added as linear constraints to \eqref{optim}, such as the maximum RWA of a legal entity is \$10 billion; or the ratio of RWA/LBS of a business unit has to be between 3 and 5. In the most general case, \eqref{optim} can be solved numerically using quadratic programming. However, if we ignore the additional business constraints, an analytical solution to \eqref{optim} can be easily obtained using Lagrange multiplier:
\begin{equation}
\label{lo}
\vec \delta^* = (J+\epsilon V^{-1})^{-1}(\lambda \vec r - \vec w - J \vec h)
\end{equation}
where $J = \frac{\partial \vec w}{\partial \vec h}$ is the Jacobian matrix, $\lambda$ is the Lagrange multipler which can be determined using the constraint $\vec r \cdot \vec \delta = z$: 
\[
\lambda = \frac{ z + \vec r^T (J+\epsilon V^{-1})^{-1}\vec w + \vec r^T (J+\epsilon V^{-1})^{-1}J \vec h}{\vec r^T (J+\epsilon V^{-1})^{-1}\vec r}
\]
The Jacobian matrix $J$ can be easily computed numerically or analytically.

If we ignore the changes in the capital ``exchange rates'' $\vec w$ under small changes in $\vec h$ by setting $J=0$, \eqref{lo} reduces to a simpler ``crude'' approximation:
\begin{equation}
\label{rot}
\vec \delta^* = \lambda V\left(\vec r - \frac{\vec w}{\lambda}\right)
\end{equation}
where $\frac{\vec w}{\lambda}$ is a RoC threshold.

The crude solution \eqref{rot} has a simple interpretation when $V$ is diagonal: it dictates the growth of the business units with high RoCs at the expense of those with low RoCs. It mimics the rule-of-thumb approach that is often used in the absence of a rigorous quantitative solution for capital optimization. The \eqref{rot} shows that the rule-of-thumb approach can only be valid under the following two conditions: 1) $V$ is diagonal, i.e., there is zero correlation between the changes in different capital components 2) the RoCs are computed using Shapley or CAS allocation, as it is a pre-requisite to set up the local optimization problem.

\begin{table}
\centering
\caption{Bank's Revenue and Capital}
\vspace{.2cm}
\label{tab2}
\small
\begin{tabular}{c|rr|r|rr|rr}
\hline
\multirow{2}{*}{\textbf{\begin{tabular}[c]{@{}c@{}}Business\\ Unit\end{tabular}}} & \multicolumn{1}{c|}{\multirow{2}{*}{\textbf{\begin{tabular}[c]{@{}c@{}}RWA\\ Capital\end{tabular}}}} & \multicolumn{1}{c|}{\multirow{2}{*}{\textbf{\begin{tabular}[c]{@{}c@{}}LBS\\ Capital\end{tabular}}}} & \multicolumn{1}{c|}{\multirow{2}{*}{\textbf{Revenue}}} & \multicolumn{1}{c|}{\multirow{2}{*}{\textbf{\begin{tabular}[c]{@{}c@{}}Return on\\ RWA/C\end{tabular}}}} & \multicolumn{1}{c|}{\multirow{2}{*}{\textbf{\begin{tabular}[c]{@{}c@{}}Return on\\ LBS/C\end{tabular}}}} & \multicolumn{2}{c}{\textbf{RoC Threshold}}                   \\ \cline{7-8} 
                                                                                  & \multicolumn{1}{c|}{}                                                                                & \multicolumn{1}{c|}{}                                                                                & \multicolumn{1}{c|}{}                                  & \multicolumn{1}{c|}{}                                                                                    & \multicolumn{1}{c|}{}                                                                                    & \multicolumn{1}{c|}{\textbf{RWA/C}} & \multicolumn{1}{c}{\textbf{LBS/C}} \\ \hline
A                                                                                 & 230                                                                                                  & 150                                                                                                  & 23                                                     & 0.0605                                                                                                   & 0.0605                                                                                                   & 0.0365                              & 0.0879                             \\
B                                                                                 & 120                                                                                                  & 250                                                                                                  & 25                                                     & 0.0676                                                                                                   & 0.0676                                                                                                   & 0.0365                              & 0.0879                             \\
C                                                                                 & 150                                                                                                  & 250                                                                                                  & 25                                                     & 0.0625                                                                                                   & 0.0625                                                                                                   & 0.0365                              & 0.0879                             \\
D                                                                                 & 250                                                                                                  & 150                                                                                                  & 25                                                     & 0.0625                                                                                                   & 0.0625                                                                                                   & 0.0365                              & 0.0879                             \\
E                                                                                 & 150                                                                                                  & 200                                                                                                  & 20                                                     & 0.0571                                                                                                   & 0.0571                                                                                                   & 0.0365                              & 0.0879                             \\ \hline
\textbf{Total}                                                                    & 900                                                                                                  & 1000                                                                                                 & 118                                                    &                                                                                                          &                                                                                                          &                                     &                                   
\end{tabular}
\end{table}

\begin{table}[]
\centering
\caption{\label{tab3} Local Capital Optimization}
\small

\vspace{.5cm}
\underline{Local Optimum $\vec \delta^*$ with $z=0, \epsilon=0.1, V=I$}
\vspace{.2cm}

\begin{tabular}{c|rrr|rrr}
\hline
\multirow{2}{*}{\textbf{\begin{tabular}[c]{@{}c@{}}Business\\ Unit\end{tabular}}} & \multicolumn{3}{c|}{\textbf{Local Optimal (11)}}                                                                & \multicolumn{3}{c}{\textbf{Crude Solution (12)}}                                                               \\ \cline{2-7} 
                                                                                  & \multicolumn{1}{c|}{\textbf{RWA/C}} & \multicolumn{1}{c|}{\textbf{LBS/C}} & \multicolumn{1}{c|}{\textbf{Total}} & \multicolumn{1}{c|}{\textbf{RWA/C}} & \multicolumn{1}{c|}{\textbf{LBS/C}} & \multicolumn{1}{c}{\textbf{Total}} \\ \hline
A                                                                                 & 1.45                                & -1.73                               & 1.09                                & 1.49                                & -1.70                               & 1.18                               \\
B                                                                                 & 2.03                                & -1.15                               & -4.23                               & 1.93                                & -1.26                               & -4.41                              \\
C                                                                                 & 1.61                                & -1.57                               & -3.85                               & 1.61                                & -1.58                               & -3.91                              \\
D                                                                                 & 1.61                                & -1.57                               & 1.80                                & 1.61                                & -1.58                               & 1.85                               \\
E                                                                                 & 1.17                                & -2.01                               & -2.84                               & 1.28                                & -1.91                               & -2.75                              \\ \hline
\textbf{Total}                                                                    & 7.88                                & -8.03                               & -8.03                               & 7.92                                & -8.03                               & -8.03                             
\end{tabular}

\vspace{.5cm}
\underline{Local Optimum $\vec \delta^*$ with $z=0$, Corr(dLBS, dRWA)=0.95}
\vspace{.2cm}

\begin{tabular}{c|rrr|rrr}
\hline
\multirow{2}{*}{\textbf{\begin{tabular}[c]{@{}c@{}}Business\\ Unit\end{tabular}}} & \multicolumn{3}{c|}{\textbf{Local Optimal (11)}}                                                                & \multicolumn{3}{c}{\textbf{Crude Solution (12)}}                                                               \\ \cline{2-7} 
                                                                                  & \multicolumn{1}{c|}{\textbf{RWA/C}} & \multicolumn{1}{c|}{\textbf{LBS/C}} & \multicolumn{1}{c|}{\textbf{Total}} & \multicolumn{1}{c|}{\textbf{RWA/C}} & \multicolumn{1}{c|}{\textbf{LBS/C}} & \multicolumn{1}{c}{\textbf{Total}} \\ \hline
A                                                                                 & -2.70                               & -5.15                               & -3.10                               & -2.05                               & -4.62                               & -2.42                              \\
B                                                                                 & 13.49                               & 11.04                               & 9.05                                & 11.72                               & 9.14                                & 6.94                               \\
C                                                                                 & 1.84                                & -0.61                               & -2.31                               & 1.81                                & -0.76                               & -2.58                              \\
D                                                                                 & 1.84                                & -0.61                               & 1.98                                & 1.81                                & -0.76                               & 1.98                               \\
E                                                                                 & -10.48                              & -12.93                              & -13.87                              & -8.66                               & -11.24                              & -12.16                             \\ \hline
\textbf{Total}                                                                    & 4.00                                & -8.25                               & -8.25                               & 4.64                                & -8.25                               & -8.25                             
\end{tabular}

\vspace{.5cm}
\underline{Local Optimum $\vec \delta^*$ with $z=2$, Corr(dLBS, dRWA)=0.95}
\vspace{.2cm}

\begin{tabular}{c|rrr|rrr}
\hline
\multirow{2}{*}{\textbf{\begin{tabular}[c]{@{}c@{}}Business\\ Unit\end{tabular}}} & \multicolumn{3}{c|}{\textbf{Local Optimal (11)}}                                                                & \multicolumn{3}{c}{\textbf{Crude Solution (12)}}                                                               \\ \cline{2-7} 
                                                                                  & \multicolumn{1}{c|}{\textbf{RWA/C}} & \multicolumn{1}{c|}{\textbf{LBS/C}} & \multicolumn{1}{c|}{\textbf{Total}} & \multicolumn{1}{c|}{\textbf{RWA/C}} & \multicolumn{1}{c|}{\textbf{LBS/C}} & \multicolumn{1}{c}{\textbf{Total}} \\ \hline
A                                                                                 & 0.44                                & -2.01                               & 0.05                                & 0.89                                & -2.86                               & 0.35                               \\
B                                                                                 & 16.98                               & 14.53                               & 12.54                               & 21.39                               & 17.64                               & 14.43                              \\
C                                                                                 & 5.08                                & 2.63                                & 0.91                                & 6.64                                & 2.88                                & 0.22                               \\
D                                                                                 & 5.08                                & 2.63                                & 5.22                                & 6.64                                & 2.88                                & 6.89                               \\
E                                                                                 & -7.51                               & -9.95                               & -10.89                              & -8.96                               & -12.71                              & -14.06                             \\ \hline
\textbf{Total}                                                                    & 20.07                               & 7.83                                & 7.83                                & 26.60                               & 7.83                                & 7.83                              
\end{tabular}
\end{table}

Table \ref{tab3} shows the results of \eqref{lo} and \eqref{rot} for the data in Table \ref{tab2}, with different $z, \epsilon$ and $V$, and a simple revenue model assuming that the return on LBS capital and RWA capital are identical for the same business unit. The column ``RWA/C'' and ``LBS/C'' are the optimal changes in the allocated RWA and LBS capital, i.e, $\vec \delta^*$; the Total column is the change in allocated capital $\vec \alpha$ for the business unit after applying the optimal change $\vec \delta^*$. The optimization results in Table \ref{tab3} are quite sensible and intuitive: the bank's total LBS and RWA capital are more balanced; when the revenue is kept constant with $z=0$, the optimization reduces the bank's overall capital; when a revenue increase of $z=2$ is required, the bank has to increase its capital. Table \ref{tab3} also shows that the crude solution produces similar results to the full solution \eqref{lo}.

Table \ref{tab3} also verifies the scope where the rule-of-thumb may apply. When $V=I$, the rule-of-thumb approach correctly identifies the general optimal direction of change in capital, which is to increase capital allocation in A and D, the business units with the greatest overall RoC under Shapley allocation (see Table \ref{tab1}). However, when $V$ implies a 95\% correlation between the changes in RWA and LBS of the same business unit, the rule-of-thumb approach fails to identify the optimal direction of change, which is to increase the capital allocation in B at the expense of E. It is worth noting that in this case, the large LBS and RWA changes of the same business unit are all to the same direction in the optimal solution, due to the preference for plausible solutions. This result shows that the rule-of-thumb approach is inadequate in practice.

The solution \eqref{lo} may break down for large $\vec \delta^*$ because its underlying assumptions, such as constant RoC $\vec r$ and Jacobian matrix $J$, may fail with large deviation from the current capital position. Nonetheless, the analytical local optimal solution of \eqref{lo} offers a valuable quantitative framework for a bank's day to day capital management, where small changes in capital position are much more common and relevant. 

\section{Conclusion}
The accurate linear approximation to the Shapley/CAS allocation of cost function of (nested) max function is an important and useful result. It establishes a single set of ``exchange rates'' between the capital components of all legal entities, making it easy for a bank's management to reason about and actively manage their RWA and LBS capital.

By leveraging this linear approximation, we formulated banks' capital optimization problem as a simple mean/variance optimization and obtained an analytical local optimal solution that maximizes the bank's overall RoC; it is a tractable and practical quantitative framework for bank's day to day capital management.

Many well-established classic portfolio optimization techniques and concepts, such as efficient frontier, can be applied to capital optimization following this reduced form formulation, greatly expanding the available toolkit for capital management.

\bibliographystyle{apalike}
\bibliography{rwa}

\newpage

\appendix
\section{\label{appa}Derivation of $p(a<b; k)$}
We consider the following two random variables for a random permutation $\tilde{\pi}$ of $n$ units:
\begin{eqnarray}
\tilde a &=& \sum_{i=1}^n \ind_i a_i \\
\tilde b &=& \sum_{i=1}^n \ind_i b_j.
\end{eqnarray}
The $\ind_i$ is the indicators of whether the unit $i$ is ahead of a given unit $k$ in a random permutation. These indicators are not independent. One can easily prove that the $\cov (\ind_i, \ind_j) = \frac{1}{12}$ for any $k$, by noticing:
\begin{eqnarray}
\mathbb E[\ind_i] &=& \mathbb E[\ind_j] = \frac{1}{2} \\
\mathbb E[\ind_i \ind_j] &=& \frac{1}{n+1} \sum_{m=0}^n \frac{m(m-1)}{n(n-1)} = \frac{1}{3}
\end{eqnarray}
where $\frac{m(m-1)}{n(n-1)}$ is the expectation of $\ind_i \ind_j$ conditioned on there are $m$ units ahead of $k$, and $\frac{1}{n+1}$ is because there is equal probability for there are 0, 1, ..., $n$ units ahead of $k$. Therefore $\cov(\ind_i, \ind_j) = \mathbb E[\ind_i \ind_j] - \mathbb E[\ind_i] \mathbb E[\ind_j] = \frac{1}{12}$, by noting $\mathbb E[\ind_i] = \frac{1}{2}$. Therefore the correlation between $\ind_i$ and $\ind_j$ is $\frac{1}{3}$ by noting the variance of $\ind_i$ is $\frac{1}{4}$.

Therefore we have: $\mathbb E[\tilde a] = \frac{1}{2} \sum_i a_i$, $\mathbb E[\tilde b] = \frac{1}{2} \sum_i b_i$ and the covariance:
\begin{eqnarray}
\label{cov}
\nonumber
\cov(\tilde a, \tilde b) &=& \sum_{i, j} \cov(\ind_i a_i, \ind_j b_j) = \sum_{i, j} a_i b_j \; \cov(\ind_i, \ind_j) \\ \nonumber
 &=& \frac{1}{4} \sum_i a_i b_i + \frac{1}{12} \sum_{i \ne j} a_i b_j \\
 &=& \frac{1}{6}\sum_i a_i b_i + \frac{1}{12} (\sum_i a_i) (\sum_j b_j)
\end{eqnarray}
The variance of $\tilde a$ is therefore $\cov(\tilde a, \tilde a)$.

\end{document}